\documentclass[10pt,a4paper]{article}
\usepackage[utf8]{inputenc}
\usepackage{amsmath}
\usepackage{amsfonts}
\usepackage{amssymb}
\usepackage{graphicx}
\usepackage{xcolor}
\usepackage{siunitx}
\graphicspath{{figures/}}

\providecommand{\keywords}[1]
{
  \small	
  \textbf{\textit{Keywords---}} #1
}

\author{
Nicola Catenacci Volpi,\and 
Martin Greaves, \and
Dari Trendafilov, \and 
Christoph Salge, \and
Giovanni Pezzulo \and
and Daniel Polani}

\title{Skilled motor control implies a low entropy of states but a high entropy of actions} 
\date{}
\begin{document}
\maketitle

\begin{abstract}
The mastery of skills such as playing tennis or balancing an inverted pendulum implies a very accurate control of movements to achieve the task goals. Traditional accounts of skilled action control that focus on either routinization or perceptual control make opposite predictions about the ways we achieve mastery. The notion of routinization emphasizes the decrease of the variance of our actions, whereas the notion of perceptual control emphasizes the decrease of the variance of the states we visit, but not of the actions we execute. Here, we studied how participants managed control tasks of varying levels of complexity, which consisted in controlling inverted pendulums of different lengths. We used information-theoretic measures to compare the predictions of alternative theoretic accounts that focus on routinization and perceptual control, respectively. Our results indicate that the successful performance of the control task strongly correlates with the decrease of state variability and the increase of action variability. As postulated by perceptual control theory, the mastery of skills consists in achieving stable control goals by flexible means. 
\end{abstract}

\section{Introduction}
 
Humans are able to learn sophisticated skills, such as playing a musical instrument professionally, excelling in complex sports such as tennis, balancing an inverted pendulum, or driving a car in a busy city. Yet, the informational principles underlying human skilled action control are incompletely known \cite{fitts1964information, crossman1964information, mackenzie1989note, koechlin2007information, haykin2012cognitive,  zenon2019information}. 

Classical theories of skill learning in motor control assume that with training, actions become routinized, which implies a drastic reduction of their variability \cite{shmuelof2012motor}. Action variability might be initially useful during learning to promote exploration and remains also afterwards, but in general, a hallmark of expertise and routinization is the low variability of skilled actions \cite{Dhawale2017,Herzfeld2014,Newell1993,Summers2009,Schmidt2018}. The idea that with expertise action variability decreases is also implicit in standard machine learning and AI approaches to gaming (e.g., deep RL systems), where the goal is to learn a convenient policy, or state-action mapping \cite{sutton2018reinforcement}. The massive learning process of deep RL systems implies a routinization of actions and the drastic reduction of the variability of the selected policies -- in the sense that at the end of learning, a single policy is retained from a potentially huge space of possible policies considered before learning. 

On the contrary, \emph{perceptual control theory} assumes that we control (and reduce the variability of) perceptual states, not actions \cite{Powers1973}. For example, when we drive a car, we strive to keep the speed indicator fixed on our desired speed (say, 80 mph). Therefore, we allow for lower variability on a perceptual variable (speed indicator) but higher variability on our actions (press or release break) to achieve our control objective. Other related frameworks, such as planning-as-inference, active inference, surprise-minimising RL and KL control are also based on the idea that goal-directed behavior amounts to reducing the entropy or variance of the final (goal) state(s) of the controlled dynamical system  \cite{Attias2003,Botvinick2012,Friston2010, berseth2019smirl, rhinehart2021intrinsic, Kappen2012}. 
For example, when balancing an inverted pendulum, the task ends in one single state, which is the one with the pendulum up and still. In most practical cases, it is important to reduce the entropy not just of the final (goal) state, but also of (some of) the intermediate states that the agent has to visit to achieve its goals. This is because in difficult control tasks, such as balancing an inverted pendulum, there is only a ``narrow corridor" in the phase space of the task: only a very small subset of trajectories (compared to the full set of possible trajectories in the state space) afford goal achievement. In other words, a difficult control task implies that only a small region of the state space must be visited with high probability to reach the goal. Some exploration of the state space (here, indexed by high state entropy) might be useful during initial phases of learning, or when the control task is easy, but could otherwise hinder the performance.

We designed an experiment to adjudicate between these contrasting views of expertise -- and to assess the relative importance of reducing variability of actions and states during skilled control. We analysed human participants' behaviour and performance during a series of control tasks, at various levels of difficulty.  The task consisted in balancing an inverted pendulum during different trials with varying pendulum length \cite{lupu2014information, loram2011human}, which corresponds to the experimental condition used to study control tasks of different difficulty. We reasoned that, if the former hypothesis (reduction of action variability) is correct, skilled control of a continuous dynamical system should be indexed by low levels of action entropy. Rather, if the second hypothesis (reduction of perceptual state or outcome variability) is correct, skilled control should be indexed by low levels of state entropy, but not action entropy.

\section{The experiment}

We studied the behaviour of 36 human participants performing a continuous control task, consisting in inverting and balancing a  pendulum about its axis for as long as possible, using a computer simulation of a swinging pendulum. 

The participants of the experiment were adults of mixed age and gender   recruited from the students and staff of the University of Hertfordshire. Subjects had no prior knowledge of the project or the specific task being employed for testing. The experiment was approved by the Ethics Committee of the University of Hertfordshire as complying with the University's policies regarding studies involving the use of human participants (study approved by University of Hertfordshire Science and Technology Ethics Committee and Designated Authority with Protocol No.  COM SF UH 00016). Informed consent was provided by all participants.

\subsection{Controlled System}

\begin{figure}[h]
  \centering
    \includegraphics[width=0.4\columnwidth]{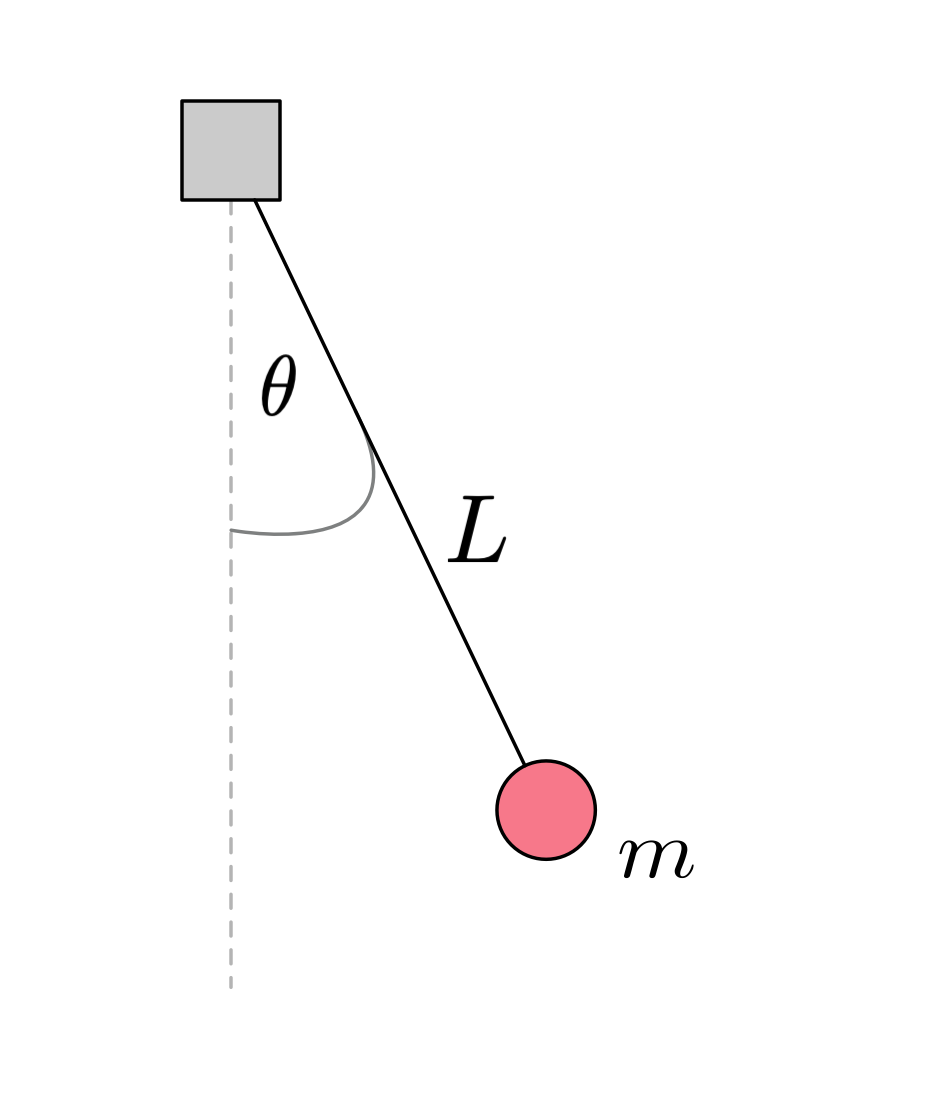}
    \caption{The controlled system is a pendulum with mass $m$ hanging from a weightless rod of length $L$. The angle between the pendulum and its vertical axis is denoted by $\theta$, where $\theta =$ \SI{0}{\radian} indicates the pendulum's downward position. The goal of the task assigned to participants is to swing the pendulum in order to balance it in its upright position for as long as possible (i.e., $\theta \simeq$ \SI{\pi}{\radian} and $\dot{\theta} \ll$ \SI{1}{\radian\per\second}).}
    \label{fig:pendulum}
\end{figure}

The non-linear dynamical system controlled by the participants is a pendulum composed by a mass $m$ hanging from a weightless rod of length $L$ (see Figure \ref{fig:pendulum}).  We denote with $\theta_t$ the angle that the pendulum has with its vertical axis at time $t$, with $\theta_t =$ \SI{0}{\radian} indicating the downward position. In the experiment we set $m= $ \SI{1}{\kilogram} and $g=$ \SI{1}{\meter \per \square \second} for the gravitational acceleration $g$.  Hence, the dynamics of the controlled pendulum is governed by the following differential equation

\begin{equation}
\label{eq:pendulum_ac}
\ddot{\theta}_t \doteq \frac{d^2 \theta_t}{dt^2} = \frac{u_t - \sin \theta_t}{L}
\end{equation}
\\
where $u_t$ represents the contribution of the mass acceleration controlled by the participants at time $t$.  In the simulation this equation is approximated using the Euler method for numerical integration.
Given a time interval $\Delta t$,  in the numerical simulation the pendulum's angular speed $\dot{\theta}_{t+\Delta t}$ is computed as follows

\begin{equation}
\label{eq:pendulum_vel}
\dot{\theta}_{t+\Delta t}  = (1 - d) \dot{\theta}_t + \ddot{\theta}_t \Delta t
\end{equation}
\\
where $d$ is a damping coefficient set to $5 \times 10^{-6}$ in our experiments.  Pendulums with different lengths $L$ will exhibit different periods, with larger values of $L$ implying longer periods. Hence,  a shorter pendulum manifests larger angular velocities  $\dot{\theta}$ than a longer pendulum,  making it more difficult to control.

\subsection{Control Task}

In the experiment, participants were required to balance a pendulum vertically above its axis in a computer simulation and keep it upright for as long as possible.  Participants controlled the pendulum using two keys of the computer's keyboard,  corresponding to the controlled external mass accelerations $u = \pm$ \SI{0.27}{\meter \per \square \second} respectively.  To hold down a key  exerts a continuous force until the key is released (i.e., the subject did not need to repeatedly press a key to exert a continuous force but just to hold it down). Note that both keys can act as brake or accelerator according to the direction of the pendulum.

The structure of the experiment was as follows. Prior to starting the experiment, an instructor demonstrated to the participants what is required to perform the task. Then, the participants performed 4 practice trials in which they learned how to control and balance the pendulum. The time allowed to balance the pendulum on each training trial was 3 minutes. In these practice trials, the length of the pendulum $L$ was varied, to let participant experience tasks of different levels of difficulty: easier for longer pendulums (having smaller angular speed) and more difficult for shorter pendulums (having larger angular speed). 

The main experiment consisted of 8 trials lasting 2 minutes each, in which participants were asked to balance 8 different pendulum lengths $L$ (i.e., difficulty levels). The values of $L$ are: $L=0.2,0.3,0.4,0.5,0.6,0.7,0.8$ and  \SI{0.9}{\metre}, with $L=$ \SI{0.2}{\metre} being the most difficult control task and $L=$ \SI{0.9}{\metre} the easiest one. The experimental condition changed according to a within-subjects design, so all participants were tested for all difficulty levels $L$ with the order of trials being randomised across participants. By manipulating the pendulum length $L$ (and hence the speed at which the pendulum swings around its axis) we controlled the difficulty of the task. This allowed us to investigate the control strategies and the possible reduction of action and/or state variability with respect to the limits of human processing capacity.

More formally, the objective of each trial is to swing the pendulum from a starting configuration where it is still and downward ($\theta=$ \SI{0}{\radian} and $\dot{\theta} =$ \SI{0}{\radian\per\second}) in order to reach the upright configuration ($\theta =$ \SI{\pi}{\radian}). Then, participants were required to use the keys to keep the pendulum balanced in this upright position for as long as possible (i.e., $\theta \simeq$ \SI{\pi}{\radian} and $\dot{\theta} \ll$ \SI{1}{\radian\per\second}). We emphasise that in principle it is possible to design am artificial controller capable of balancing the pendulum of different lengths using the two provided mass accelerations $u$. However, in practice, when the task is addressed by humans, their information processing capacity is limited and therefore controlling shorter pendulums becomes increasingly hard.

Participants were informed that balancing the pendulums (especially the shorter ones) was challenging. Subjects were told that failure to balance may be a possibility but were instructed to do their best in each case.  Furthermore,  subjects were informed that the times for which the pendulum was balanced on its axis was measured.

\section{Methods}

Let us denote by $s_t \doteq (\theta_t,\dot{\theta}_t) \in  \mathcal{S} \doteq [0, 2 \pi) \times \mathbb{R}$ the pendulum's \emph{state} at time $t$, being a two-dimensional tuple composed by the pendulum's angle $\theta$ and angular speed $\dot{\theta}$.  In addition, let us denote with $a_t \in  \mathcal{A} \doteq \{l,n,r\}$ the \emph{action} that a participant selects at time $t$,  being either to press the key that increases the pendulum's swing in the clockwise direction (``left" action $a_t = l$), to press the key that increases the pendulum's swing in the anti-clockwise direction (``right" action $a_t = r$), or simply to do nothing (``no-action" $a_t = n$).

For each trial $i$, we collected the states $\overrightarrow{S}^i \doteq (s_0^i, s_{\Delta t}^i, s_{2 \Delta t}^i, \dots, s_{t_F}^i)$ visited by the pendulum and the actions $\overrightarrow{A}^i \doteq (a_0^i, a_{\Delta t}^i, a_{2 \Delta t}^i, \dots, a_{t_F}^i)$ chosen by the participant,  where $\Delta t$ denotes the sample interval and ${t_F}$ the trial's final time. In the experiment we set $\Delta t \simeq$ \SI{0.017}{\second} and ${t_F} =$ \SI{120}{\second}, hence each trial $i$ is represented by the two time series $\overrightarrow{S}^i$ and $\overrightarrow{A}^i$, each one composed of $F=7161$ samples.

To investigate the relationship between skilled control and the variability of actions and states, we analysed the aforementioned time series measuring the performance and estimating information-theoretic measures of variability for each participant in each experimental condition. To measure participants' performance we employed a mathematical framework commonly used in models of sequential decision-making and control \cite{bertsekas2000dynamic, sutton2018reinforcement,dayan2001theoretical}. We first defined the \emph{reward} as a function of the state space $r: \mathcal{S} \rightarrow \{0,c\}$,  which measure the immediate worth of a state with respect to the given balancing task.  In the experiment we defined the reward as follows

\begin{equation}
\label{eq:reward}
r(s_t) =  \left\{\begin{array}{cl}  0  & \mbox{if \;  $ ( \theta_t \in (\pi - 0.26, \pi + 0.26)\, \SI{}{\radian}) \; \; and \; \; (\mid \dot{\theta}_t \mid < \SI{0.2}{\radian\per\second})$} \\
c \in  \mathbb{R}^- & \mbox{otherwise} \\
\end{array}\right.
\end{equation} 

In other words, $r(s_t)$ penalises states that are not close to a balanced configuration, which are states where the pendulum is far from its upright position and the angular speed is far from zero.  In the experiment we chose $c = -3 \times 10^{-3}$. Then, in order to evaluate an entire trial, we defined the \emph{utility} of a trial $i$, denoted as $U^i$, as the sum of all rewards accumulated by the participant for all time steps of the trial. Hence, for trial $i$ we have

\begin{equation}
\label{eq:reward}
U^i \doteq \sum_{n=0}^F r(s^i_{n \Delta t})
\end{equation} 

We quantify actions' and states' variability employing the information-theoretic notion of \emph{entropy}, which is used to measure the uncertainty of a random variable \cite{cover1999elements}. Let us denote by $A^i \in \mathcal{A}$ the random variable representing the identically distributed actions chosen at every time step $t$ by the subject participating in the trial $i$. Let us also denote by $S^i\in \bar{\mathcal{S}}$ the random variable representing the identically distributed states visited at every time step $t$ by the pendulum during the trial $i$, where $\bar{\mathcal{S}}$ is a discretised version of $\mathcal{S}$ obtained through binning. Let us also assume that the time series $\overrightarrow{A}^i$ and $\overrightarrow{S}^i$ are realizations of sequential samplings of the random variables $A^i$ and $S^i$, respectively.    
Then,  the entropy of $S^i$ is defined as

\begin{equation}
\label{eq:entropy}
H(S^i) \doteq - \sum_{s^i \in  \bar{\mathcal{S}}} P(s^i) \log P(s^i)
\end{equation}
\\
where we denoted by $P(s^i)$ the probability of the random variable $S^i$ being equal to the state $s^i$ (i.e., $Pr\{S^i = s^i\}$). The \emph{conditional entropy} $H(A^i|S^i)$ quantifies the uncertainty left about the action $A^i$ once the state $S^i$ is known. It is defined as follows

\begin{equation}
\label{eq:cond_entropy}
H(A^i|S^i) \doteq - \sum_{s^i \in \bar{\mathcal{S}}} P(s^i) \sum_{a^i \in  \mathcal{A}} P(a^i|s^i) \log P(a^i|s^i)
\end{equation}
\\
$H(S^i|A^i)$ is defined similarly. The reduction of uncertainty about $S^i$ once the variable $A^i$ is known (and vice versa) can be measured by the \emph{mutual information} $I(S^i;A^i)$. This also quantifies in bits the amount of information that $S^i$ and $A^i$ have in common. The mutual information between states and actions is defined as

\begin{equation}
\label{eq:mutual_information}
I(S^i;A^i) \doteq \sum_{s^i \in \bar{\mathcal{S}}} \sum_{a^i \in  \mathcal{A}} P(s^i,a^i) \log \frac{P(s^i,a^i)}{P(s^i)P(a^i)}
\end{equation}
\\
The mutual information can also be rewritten as $I(S^i;A^i) = H(S^i) - H(S^i|A^i) = H(A^i) - H(A^i|S^i)$, i.e., it is symmetric.

\section{Results}

Here, we used information-theoretic measures to investigate the variability of the states $\overrightarrow{S}$ visited by the controlled system and the actions $\overrightarrow{A}$ chosen by the participants to control it.  Specifically, we analysed the quantities $H(S^i)$, $H(A^i|S^i)$ and $H(S^i|A^i)$ and $I(A^i;S^i)$ for all trials $i$. While different methods are available to estimate the above quantities from time series data collected during an experiment, we adopted the numerical estimator MIToolbox \cite{brown2012conditional}, which employs conditional likelihood maximisation to compute the entropy.  To use this discrete random variables estimator we discretized the state data collected during the experiment, binning the angle interval $[0,  2 \pi)\, $\SI{}{\radian} into 360 bins of equal size and the angular speed interval $[-10,  10]\, $\SI{}{\radian\per\second} into 200 bins of equal size. For correlation analyses, we used IBM SPSS Statistics (Version 25.0).

\subsection{Direct relation between utility and action entropy in skilled action control}

\begin{figure}[h]
  \centering
    \includegraphics[width=1\columnwidth]{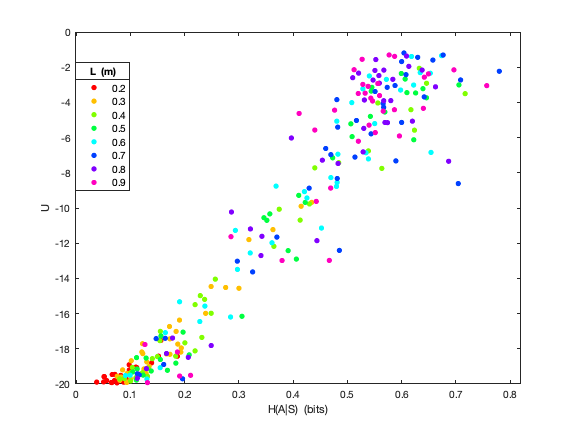}
    \caption{Utility $U^i$ of trials $i$ during the balancing task plotted as a function of $H(A^i|S^i)$ for all participants and pendulum lengths $L$.  Larger values of utility $U^i$ correspond to better performance in balancing the pendulum.}
    \label{fig:H(A|S)}
\end{figure}

The first question we addressed is whether, in keeping with classical theories of action routinization, skilled control is indexed by low levels of action entropy, given the states where the actions are executed, i.e., $H(A^i|S^i)$. For this, we investigated how the utility $U^i$ (and hence participants' skill in the pendulum balancing tasks) varies as a function of action entropy. We found a direct relation between the utility achieved by participants and the entropy of their action distributions $H(A^i|S^i)$ (see Figure \ref{fig:H(A|S)}). A significant positive Spearman correlation was observed between $H(A^i|S^i)$ and $U^i$ ($\rho = 0.944, p < 0.01$) for the whole set of trials $i$ ignoring $L$, which shows a strong monotonic relationship between the two variables (i.e., when the entropy $H(A^i|S^i)$ increases the utility $U^i$ tends to increase). In Figure \ref{fig:H(A|S)} each data point is coloured according to the difficulty level of the corresponding trial (i.e, pendulum length $L$ -- with $L= $ \SI{0.9}{\meter} being the easiest pendulum to control). As expected, utilities $U^i$ are higher for (longer) pendulums that are easier to control than for (shorter) pendulums that are more difficult to control.
Participants who controlled the pendulums better had a larger variability in their action distribution,  represented by a larger conditional action entropy $H(A^i|S^i)$, as visible in the figure. These results reject the hypothesis that skilled control is indexed by low levels of action entropy -- and indicate instead that they are indexed by high levels of action entropy. 

\subsection{Inverse relation between utility and state entropy in skilled action control}

The second question we addressed is whether, in keeping with perceptual control theory, skilled control is indexed by low levels of state entropy $H(S^i)$ and/or state action-conditioned entropy $H(S^{'i}|A^i)$. For this, we considered how utility $U^i$ varies as a function of these two entropies. By $S^{'i}$ we denote the random variable indicating the state visited by the pendulum 250 ms after an action $A^i$ has been selected. The quantity $H(S^{'i}|A^i)$ can be interpreted as the amount of uncertainty left in the joint distribution $P(A^i,S^{'i})$ about the state $S^{'i}$ once the uncertainty about the action $A^i$ has been removed. A lower $H(S^{'i}|A)$ for experts than for non-experts means that to know which action $A^i$ experts chose implies smaller uncertainty and variability about the resulting state $S^{'i}$. 

\begin{figure}[h]
  \centering
    \includegraphics[width=1\columnwidth]{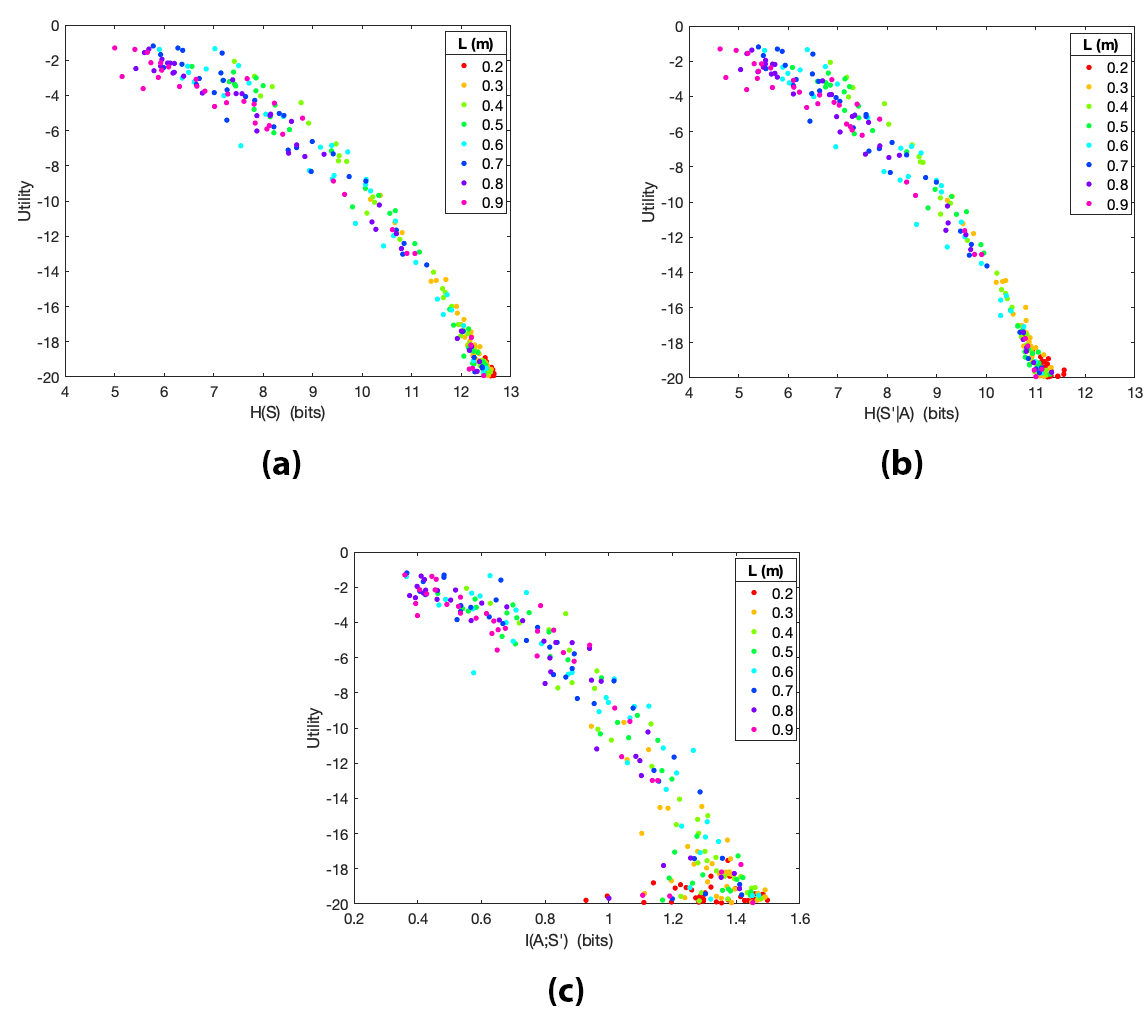}
    \caption{Utility $U^i$ of trials $i$ during the balancing task plotted as a function of $H(S^i)$ in (a), of $H(S^{'i}|A^i)$ in (b) and of $I(A^i;S^{'i})$ in (c), for all participants and pendulum lengths $L$.}
    \label{fig:entropies_information}
\end{figure}

Figures \ref{fig:entropies_information}.a and \ref{fig:entropies_information}.b show an inverse relation between utility and both the entropy $H(S^i)$ of states and the entropy $H(S^{'i}|A^i)$ of future states given participants' actions. As in the previous analysis, utilities $U^i$ are higher for pendulums that are easier to control than for pendulums that are more difficult to control.
The Spearman coefficients of the utility $U^i$ with the entropies  $H(S^i)$ ($\rho = -0.982, p < 0.01$) and $H(S^{'i}|A^i)$ ($\rho = -0.979, p < 0.01$ were computed for all trials $i$ and independently from $L$, indicating a strong correlation in the negative direction (i.e., when the state entropies decrease the utility tends to increase). These results lend support to the hypothesis that skilled control is indexed by low \emph{state} entropy. They are also compatible with the idea that expert participants perform actions that decrease the entropy of their future states. 

The two entropies $H(S^i)$ and $H(S^{'i}|A^i)$ in Figure \ref{fig:entropies_information}.a,b have a similar trend, but the latter is slightly smaller on average (their difference averaged for all trials is 1.03 bits and close to 1.5 bits for less skilled participants). This shows that most of the information contained in $H(S^{'i}|A^i)$ comes from the information contained in $H(S^i)$, but, compared to the latter, is reduced by an average difference of about 1 bit. So, although most of the information content is carried by state information, some information is absorbed (i.e., can be predicted) by the action selected by participants. This represents the amount of information the actions $A^i$ provide about the resulting states $S^{'i}$ independently from the previous states $S^i$. Note that this information is not far from $\log_2(3)$, i.e., the information that distinguishes between the three regions that the action is pushing the pendulum towards by choosing $r$, $l$ or $n$.  Since $S^{'i}$ is essentially $S^i$ shifted in time, we can assume that $H(S^i)$ is equal to $H(S^{'i})$. Hence, the aforementioned difference becomes the mutual information $I(A^i;S^{'i}) = H(S^{'i}) - H(S^{'i}|A^i)$ (see Figure \ref{fig:entropies_information}.c), which removes from the information contained in $S^{'i}$ the information contained in $S^{'i}$ knowing already $A^i$ \cite{polani2009information, genewein2015bounded, lai2021policy, butts2003much, fairhall2012information}.  Note that the results involving the mutual information $I(A^i;S^{'i})$ are consistent with those obtained with the Kraskov-St\"ogbauer-Grassberger (KSG) method for estimating the mutual information for time series with continuous data \cite{kraskov2004estimating},  which was used in addition to \cite{brown2012conditional} to verify our findings. These results are not presented here because they do not provide additional insights. 

The mutual information $I(A^i;S^{'i})$ is reported here for consistency although in character it is just reflecting the entropies. 
The fact that $I(A^i;S^{'i})$ is basically just the information injected by the action shows that the environment does not absorb the action entropy, but rather determines the subsequent behaviour. In other words, the actions chosen by the participants have a significant effect on the resulting behaviour of the pendulum. The narrow and neat character of the curves reported in Figure \ref{fig:entropies_information} is a strong indicator that their trends are general across participants rather than reflecting the particular decisions of a subset of subjects. The only case where the data points are more spread is the wide ``foot" of the mutual information curve, which shows that when the pendulum becomes very hard to control, the relationships between information and utility become fuzzier. 

The Spearman correlation of the mutual information $I(A^i;S^{'i})$ with the utility $U^i$ ($\rho =  -0.907, p < 0.01$) for all $i$ shows the similar strong negative trend found for the state entropies. Similarly to our findings, in Lupu et al. \cite{lupu2014information, lupu2013human} the average information-transmission rate of humans balancing an inverted pendulum is shown to be inversely related with the pendulum's length. The latter is used by the authors to parametrize the system's stability and the time delays experienced by the human subjects. In their work, the mutual information rate of the control systems is considered and it is estimated using the Kolmogorov-Sinai entropy, whereas in our study we estimated the Shannon mutual information of states and actions directly from data.

\begin{figure}[h!]
  \centering
  \includegraphics[width=1\columnwidth]{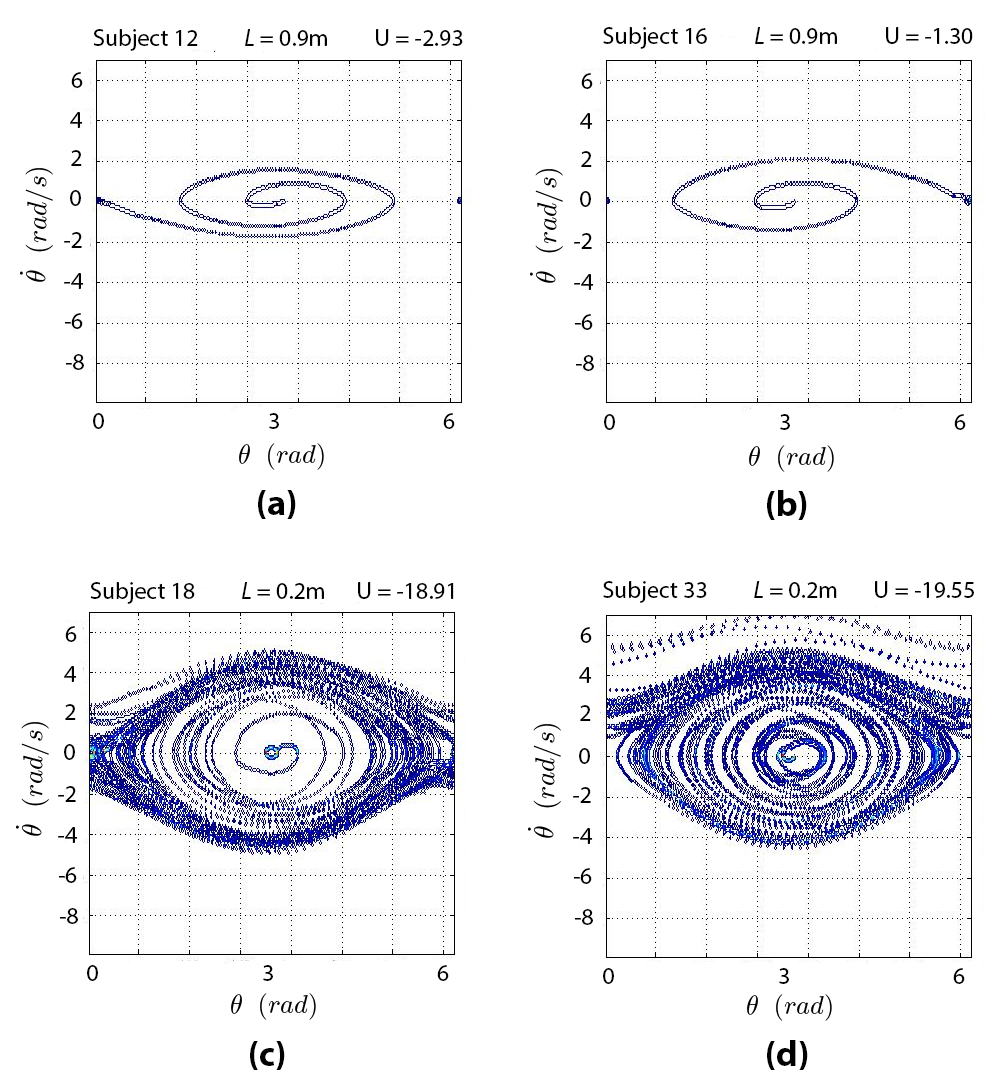}
    \caption{Support of $S^i$ within the phase space of four representative trials: two subjects swinging the longest pendulum (a,b) and other two participants swinging the shortest pendulum (c,d). Trials start at the center of the phase space.}
    \label{fig:pendulum_phase_space}
\end{figure}

To further analyze the relationship between utilities and state entropy, we plotted the support of $S^i$ over the phase space of two pendulums: the former longer and easier to control ($L=$ \SI{0.9}{\meter}, see Figure \ref{fig:pendulum_phase_space}.a,b) and the latter shorter and more difficult to control ($L=$ \SI{0.2}{\meter}, see Figure \ref{fig:pendulum_phase_space}.c,d). Our results indicate that the support of $S^i$ is smaller when the pendulum is controlled successfully -- as it is the case for most participants with a long pendulum (see Figure \ref{fig:pendulum_phase_space}.a,b for two examples). On the contrary, the support of $S^i$ is wider when the pendulum is not controlled properly -- as it is the case for most participants with a short pendulum (see for instance Figure \ref{fig:pendulum_phase_space}.c,d). This analysis further supports the idea that a hallmark of skilled action control consists in maintaining a low entropy of state distributions across the whole task, providing a visual illustration of the notion of a ``narrow corridor" in the phase space of the task mentioned in the Introduction.

\subsection{Analysis of time spent in the balancing region}

We conducted an additional analysis to test whether participants used different strategies when they are in different phases of the task: namely, when they were within or outside the balancing regions.

\begin{figure}[h]
  \centering
    \includegraphics[width=0.7\columnwidth]{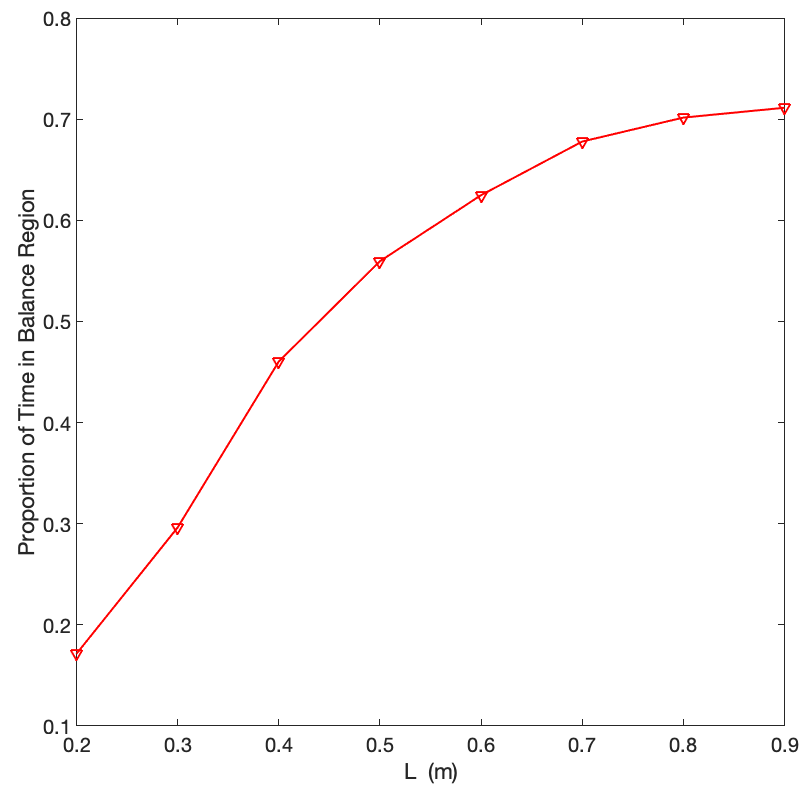}
    \caption{Proportion of the time spent by participants in the balancing region ($\pm$ 0.26 rad from the vertical axis) for each pendulum length $L$ }
    \label{fig:trial_time_balanced}
\end{figure}

\begin{figure}[h]
  \centering
  \includegraphics[width=0.7\columnwidth]{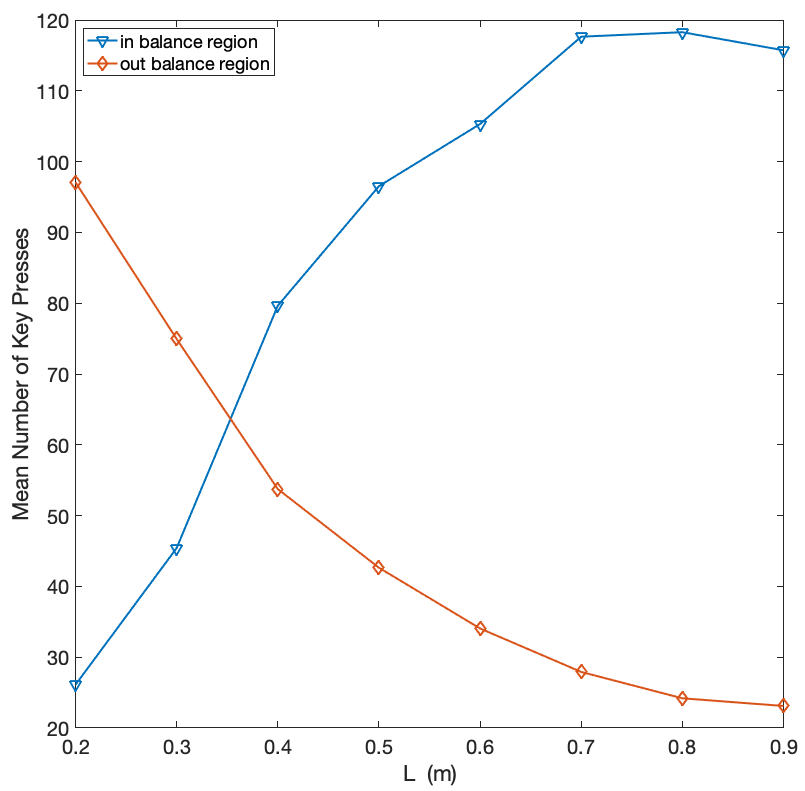}
    \caption{Key press activity within and outside balance region ($\pm$ 0.26 rad from the vertical axis) for each pendulum length $L$.}
    \label{fig:key_press_in_out_balance_region}
\end{figure}

First, as a sanity check, we tested whether participants spend less time in the balancing region when the pendulum is more difficult to control. The analysis shown in Figure \ref{fig:trial_time_balanced} confirms this prediction. Furthermore, the analysis of button presses within or outside the balancing regions indicates that participants do more key presses in the balancing region (Figure \ref{fig:key_press_in_out_balance_region}) -- as several micro-adjustments may be necessary to keep the pendulum in the correct position -- increasing action variability for skilled participants. 


\begin{figure}[h]
  \centering
    \includegraphics[width=1\columnwidth]{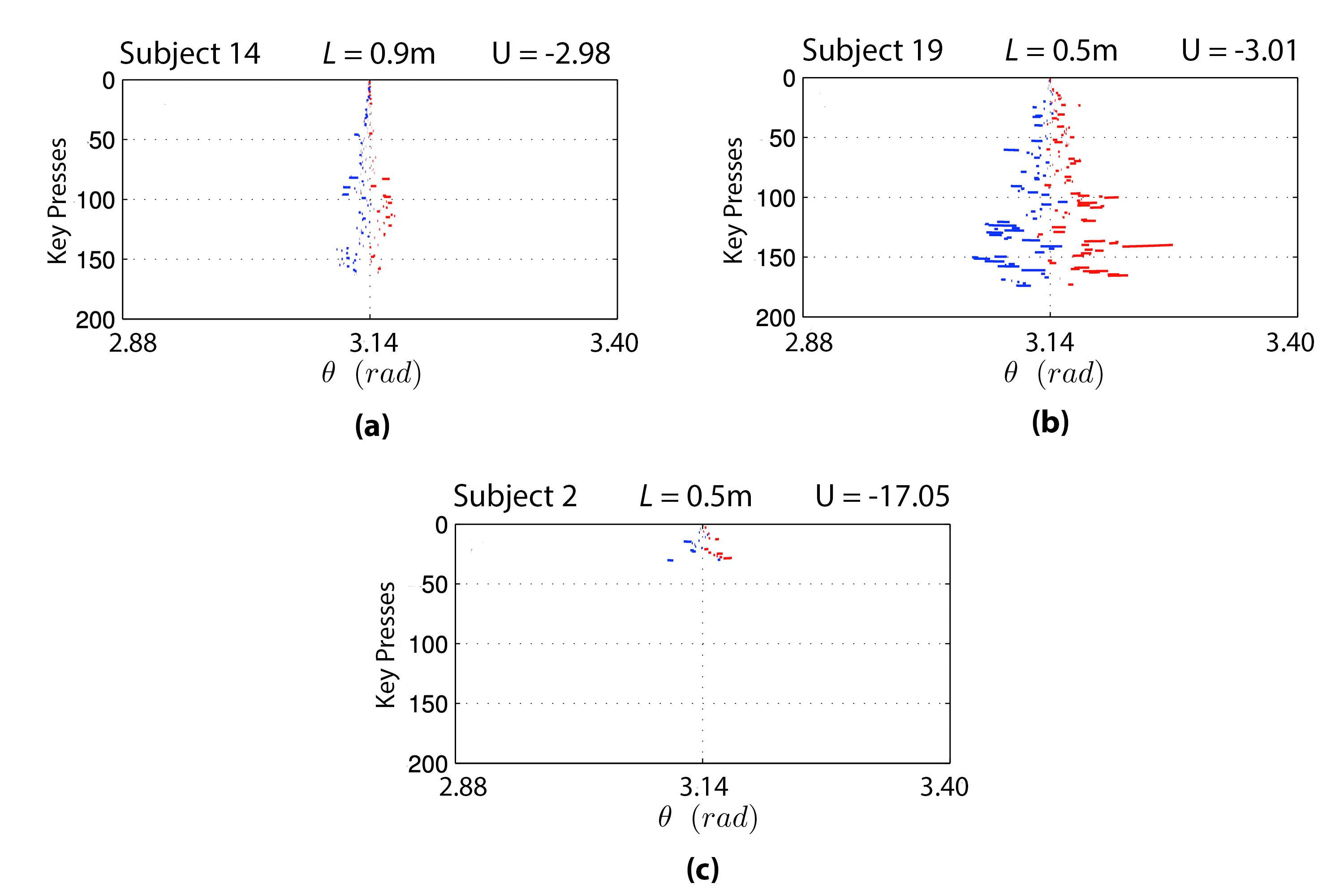}
    \caption{Distribution of key presses in the balancing region for three trials and  distance travelled by the pendulum during the key presses. Blue segments represent actions that push the pendulum in the clockwise direction, red segments indicate actions that steer the pendulum towards the anti-clockwise direction.  Key presses (``balancing episodes") are reported subsequently from the top to the bottom of the plot.  In (a,b) the key presses of two skilled participants are reported,  where in (c) the key presses of a less skilled participant is shown. }
    \label{fig:key_press_distribution}
\end{figure}

A more detailed analysis of the  actions  performed by participants in the balancing region is shown in Figure \ref{fig:key_press_distribution}.  The figure shows the distance travelled by the pendulum when a key is down, with blue segments representing actions that push the pendulum in the clockwise direction and red segments actions that push the pendulum in the anti-clockwise direction.  Subsequent key presses are reported one below another, starting from the top of the plot.  

The three plots of Figure \ref{fig:key_press_distribution} permit appreciating the differences between the key presses required to balance a pendulum that is easier to control ($L=$ \SI{0.9}{\meter}, see Figure \ref{fig:key_press_distribution}.a) versus a pendulum that is more difficult to control ($L=$ \SI{0.5}{\meter}, see Figure \ref{fig:key_press_distribution}.b,c). Furthermore, the figures show the differences between selected participants with higher skill (Figure \ref{fig:key_press_distribution}.a,b) and lower skill levels (Figure \ref{fig:key_press_distribution}.c). 

In general, controlling successfully the pendulum in the balancing region (keeping it upright) requires large variability in the choice and timing of actions (Figure \ref{fig:key_press_distribution}.a,b). This phenomenon is especially present in the case of the more skilled subject who controls the more challenging pendulum (Figure \ref{fig:key_press_distribution}.b). This finding generalises to the typical behaviour of more skilled participants who spend more time in the balancing region (also during difficult tasks), and manifest a large variability in their actions compared to less skilled participants who spend less time in the balancing region. However, less skilled participants perform fewer key presses in the balancing region (see Figure \ref{fig:key_press_distribution}.c), manifesting a smaller variability in their actions. 

\section{Discussion and Conclusions}

In this study, we asked whether low entropy of action or state distributions could be considered as signatures of mastery of a challenging task, such as controlling an inverted pendulum \cite{soukoreff2011entropy,lupu2014information, obayashi2009comparison}. Our results indicate that skilled participants keep the variability of the \emph{states} they occupy under control, whereas this variability is higher in less skilled participants (Figure \ref{fig:entropies_information}). In other words, our results indicate that skilled participants actively reduce state information to perform efficient control -- with this compression of the state space indexed by $H(S^i)$, which is lower for higher levels of utility. On the contrary, the variability of \emph{actions} during the control tasks is higher for skilled participants (Figure \ref{fig:H(A|S)}) -- which is in part due to the fact that skilled participants spend more time in the balancing regions, which require more fine-grained control. 

Our results are in keeping with perceptual control theory \cite{Powers1973} and related accounts that suggest that accurate motor control requires reducing the variability of states, including those that are outcomes of the chosen actions -- whereas this state compression can be achieved by variable means (hence corresponding to a high variability of actions). Our results are instead incompatible with theories that associate skill mastery with the routinization of action and the decrease of their variability \cite{huber2016persistence, yang2005learning}. Rather than being only important during early phases of learning due to exploration, motor variability remains high also during the successful execution of challenging control tasks \cite{Dhawale2017,Herzfeld2014,Newell1993,Summers2009,Schmidt2018}. In this regard, the classification of variability sources in advanced stages of motor learning of redundant tasks proposed in \cite{muller2009motor} suggests an interesting unifying framework.  In this classification, tasks can be characterised by a decrease of action variability when noise reduction plays a major role in successful performance or, alternatively, they can be characterised by both a decrease of state variability and large actions dispersion when action variables need to co-vary to increase invariance and accuracy in the state space.  This may explain why in the case of balancing an inverted pendulum we found that the action variability of skilled participants increases rather than decreases.  A further study in this direction is left for future investigations.

The compelling structure of the reported utility-entropy curves and the clear correlations that emerged from out analyses prompt the question of where the nature of this trade-off comes from. In general, we expect that the specific nature of the task is important -- and that the trade-offs that emerged from our analysis could be especially compelling in tasks which require reaching well defined (goal) states. For example, in the pendulum task, the final goal state distribution must have low entropy by definition (e.g., the correct position of the pendulum is almost unique). Especially in difficult tasks, there might be a very limited number of ways to achieve these goal states, forcing (skilled) participants to reduce the entropy of their trajectories to the goal states -- or in other words, forcing them to pass through a ``narrow corridor" of states to achieve the desired goals. In future work, we are going to use synthetic controllers (e.g., random or linear-quadratic-Gaussian) to investigate in more detail whether the trade-offs emerged from our analysis are due to a specific property of the task, of the decision-maker, or a combination of both.

It is possible that more than one mechanism is responsible for the state compression that we report here. Firstly,  to balance the pendulum its trajectories must be squeezed into the right part of phase space.  To obtain such entropy reduction the needed information \cite{touchette2004information} needs to be reflected in the actions. 
Then, at the level of representation, experts could abstract the state space to use less information to codify the task, encoding some portion of the state space more precisely and other portions in a more coarse manner. In the experiment, a more compact representation of the state space for more skilled participants is indicated by the number of effectively visited states. This was confirmed by our analysis, where more skilled controllers visited a smaller portion of the phase space (compare Figures \ref{fig:pendulum_phase_space}.a,b and \ref{fig:pendulum_phase_space}.c,d). To establish whether experts would be better than non-experts in appropriately coding the state space is left to future work, together with determining which resulting latent space these may use to efficiently balance the pendulum. The problem of identifying which of the pendulum's dimensions may be more relevant for the control task is addressed by the ``uncontrolled manifold'' hypothesis \cite{scholz1999uncontrolled}, which states that irrelevant dimensions usually remain uncontrolled by agents.

Behind the reduction of $H(S^i)$ and the increase of $H(A^i|S^i)$ there is the need of controlling the state at the expense of increasing action variability to perform efficient control. However, it is important to distinguish between controlled and uncontrolled variability. For the case of controlled variability, experts can actively choose to be variable (as it happened for $A^i$ in Figure \ref{fig:H(A|S)}); or they can choose to minimise the variability of the future trajectory (as it happened with $S^i$ in Figure \ref{fig:entropies_information}). There is a component of human control however that is not part of the participants intentions and that is injected in the pendulum dynamics and effectively becomes a part of uncontrolled variability of the system.  All the system's uncertainty is ultimately generated by the participants because the pendulum dynamics is deterministic and almost conservative.  So, uncertainty in control is essentially translated by the pendulum in uncertainty of its dynamics and ultimately is transformed in the uncontrolled variability of the state. But the entropy $H(S^i)$ alone does not allow us to distinguish between the controlled and uncontrolled variabilities of a state and ways to detect this difference will be investigated in future work.

\bibliography{hp}
\bibliographystyle{plain}

\end{document}